\documentclass[preprint,usegraphicx]{mn2e}
\def\etal{{\it et al.~}}

\begin{document}

\title
[Galaxy Gradients]
{The Correlation of Metallicity Gradient with
Galaxy Mass}
\author
[Forbes, S\'anchez-Bl\'azquez \& Proctor]
{Duncan A. Forbes\thanks{dforbes, psanchez, rproctor@swin.edu.au},
Patricia S\'anchez-Bl\'azquez\footnotemark[1]
\& Robert Proctor\footnotemark[1] \\
Centre for Astrophysics \& Supercomputing, Swinburne University,
  Hawthorn, VIC 3122, Australia}

\maketitle

\begin{abstract}

A number of previous studies have searched for a correlation
between radial metallicity gradients and early-type galaxy mass -- no
convincing trends have been found. 
Here we re-examine this issue with several key enhancements: using
total metallicity from studies that have broken the 
age-metallicity degeneracy, excluding galaxies with young stellar ages (i.e. those that have
experienced a recent central starburst) and using the K-band to
derive galaxy luminosities.  
We find that Coma
cluster galaxies have metallicity gradients which correlate with galaxy mass.
Furthermore, gradients have values
similar to those of monolithic collapse models.
The combination of dissipative
formation and energy injection from supernova provides a
mechanism 
for the trends with galaxy mass, however other explanations are
possible. 
Additional high quality observational data is needed to further
constrain the gas physics involved in galaxy formation.

\end{abstract}


\section{Introduction}

The nature of galaxy formation is one of the key questions in
contemporary astrophysics. Within the Cold Dark Matter (CDM) paradigm
(Blumenthal \etal 1984), two main scenarios for galaxy formation have
been recognised. They are {\it monolithic collapse} at early
epochs and {\it hierarchical merging} which can continue to later epochs.
%Although the distinction between the two scenarios is not well defined
%(Peebles 2002). 
%For example, the merging of numerous highly
%gas-rich subunits resembles the dissipative collapse of a clumpy gas
%cloud. 

%On the one hand, there is ample evidence at low redshift for
%ongoing accretion, interactions and outright mergers to support
%the concept of hierarchical galaxy assembly. However, other
%evidence strongly suggests that the bulk of stars in galaxies
%formed at early epochs. This evidence includes galaxy scaling relations
%(e.g. colour-magnitude relation, fundamental plane), extremely
%red galaxies, the age of globular clusters, cosmic star formation rates with
%redshift etc. Our current understanding of galaxy formation is that
%roughly 2/3 of stars formed at these early epochs (e.g. see review
%by Bell 2004). 

%Gaseous (dissipative) {\it collapse} models were partially
%motivated by the desire to explain the
%observed correlation of global metallicity with galaxy mass for
%early-type galaxies (Sandage 1972). 
A robust outcome of these early collapse models was
the presence of radial metallicity gradients (Larson 1974;
Carlberg 1984a,b). During the
collapse, stars form from gas that is chemically-enriched as it flows towards the
centre of the potential well, thus establishing a negative radial gradient
in which the central stars are more metal-rich than those in the outer
regions. 
Carlberg (1984b) showed that the injection of energy associated
with galactic winds (from stellar mass loss or supernova)
ultimately 
provides a pressure which drives the gas from the
galaxy. 
%thus decreasing the metallicity gradient. 
%His simulations gave 
The resultant metallicity gradients of --0.5 dex per dex in the
high mass galaxies, compared to low mass
galaxies with almost zero gradient, is a result of deeper potential wells retaining more
metals. 
More recent chemodynamical models, which include
metallicity-dependent cooling and supernova feedback, also
predict stronger metallicity gradients in more massive galaxies 
%have derived metallicity
%gradients for baryonic matter within dark matter halos 
%in a CDM cosmological context 
(Kobayashi 2004; Kawata
\& Gibson 2003; Chiosi \& Carraro 2002; Kawata 1999, 1999; Bekki
\& Shioya 1999). 
Radial gradients in colour (Franx \& Illingworth 1990),
line indices (Carollo \& Danziger 1994) and total metallicity
(Mehlert \etal 2003; hereafter M03) have been
shown to correlate with the {\it local} escape velocity or
gradients in velocity dispersion. 
This has lead to a different interpretion of metallicity
gradients by Martinelli \etal (1998) in which gradients are
established by the onset of a galactic
wind, which varies with the local depth of the potential well. 

The formation of elliptical galaxies by a major {\it merger} also
predicts metallicity gradients, however those gradients are
expected to be significantly shallower than mentioned above,
particularly for massive galaxies 
%those of a dissipative collapse 
(e.g. White 1980; Barnes 1996; Bekki \&
Shioya 1999). Thus observations of radial metallicity gradients
may offer a way of distinguishing between competing galaxy formation
scenarios.
% of {\it collapse} and {\it merging}. 

%Motivated by the Carlberg models, s
Several observational studies have examined metallicity gradients as a function
of galaxy mass.
%searched for a correlation of
%stronger metallicity gradients in more massive galaxies.
%, aspredicted by the Carlberg (1984b) collapse models.  
Most
previous work has focused on using colour (e.g. Peletier
\etal 1990) or raw 
absorption line indices (e.g. Carollo \etal 1993; Davies \etal
1993; Fisher, Franx \& Illingworth 1995) 
as a proxy for total metallicity. Carollo \etal (1993) claimed
that a trend was present for low mass galaxies, but
that gradients did not change with galaxy mass for the highest
mass systems. From a literature compilation of 80 early-type
galaxies, Kobayashi \& Arimoto (1999) derived metallicity
gradients from absorption line indices finding 
a typical gradient of 
%$\Delta log Z / log r$ $\sim$
--0.3 dex per dex with an rms dispersion of about 0.15. 
They found no convincing trend of metallicity gradient with galaxy mass. 

Both colour and raw line indices are affected by the
age-metallicity degeneracy. So, in a given galaxy, they trace some (unkown)
combination of age and metallicity. In order 
to disintangle these effects, and hence break the 
age-metallicity degeneracy one requires a method such as the use
of Lick line indices and single 
stellar population models (e.g. Proctor 2002; M03; 
S\'anchez-Bl\'azquez 2004; hereafter SB04) to derive independent metallicity
and age gradients. (An alternative approach to break the
degeneracy is to examine
galaxies at higher redshift, e.g. Tamura \etal 2000.) 
In his thesis, Proctor (2002) found no obvious trend with
velocity dispersion.  M03 
%study of 35 early-type galaxies in the
%Coma cluster  
%a mean metallicity
%gradient for their sample of $\Delta log Z / log r$ $\sim$
%--0.16. They 
did not find any ``significant evidence for correlations''
between total metallicity gradient and galaxy mass in their study
of 35 Coma galaxies. In her thesis, SB04 found a
weak trend with velocity dispersion for Coma cluster galaxies but only when using the combination
of H$\beta$ and Fe4383 lines.

As measured gradients are luminosity-weighted and the
central galaxy regions will be the focus of any merger-induced
starburst (Hernquist \& Barnes 1992), we would expect metallicity
gradients to be artifically enhanced in the first few Gyrs
following a central starburst.
One example of this might be the field elliptical NGC 821. 
%with a strong gradient is NGC 821. 
It has a central luminosity-weighted age of $\sim$ 4 Gyrs
indicating new star formation took place 
a few Gyrs ago, presumably induced by the accretion of
gas into the galaxy central regions. It has a very steep metallicity gradient
of $\sim$ --0.8 dex per dex (Proctor \etal 2005). 
%, i.e. a very steep gradient given its
%moderate mass (M$_B$ = --20.24, $\sigma$ = 210 km/s). 
%In the Kobayashi (2004) framework, 
%subsequent mergers, which have reduced gas fractions, tend to make the
%gradients much shallower. This is supported by the models of Bekki \&
%Shioya (1999), who found very shallow gradients in old merger remnant
%ellipticals. 
Thus previous observational studies which included 
galaxies with young central stellar populations    
(i.e. those that may have experienced a recent merger-induced
central starburst), may therefore have introduced
some scatter into any gradient versus mass correlation.

%This
%is due in part to the fact that metallicity gradients are
%luminosity-weighted and the central galaxy regions will be site
%of any merger-induced starburst.  About 5 Gyrs after the
%merger, the measured gradient is much shallower. 

Here we revisit the important issue of early-type galaxy
metallicity gradients. We use metallicity 
gradients derived from single stellar population models which break
the age-metallicity degeneracy. We also exclude galaxies with 
measured young central ages, and use the K-band to derive galaxy
luminosities.  
With these
enhancements 
we find statistically significant correlations of metallicity gradients
with galaxy mass for Coma cluster early-type galaxies. 

\section{Recent Model Predictions}

%In this section we briefly review models in the literature
%that make predictions about the variation of metallicity gradients
%with elliptical galaxy mass.

Kawata (1999) simulated the formation of a slowly rotating elliptical
galaxy modelling the gas, stars and dark matter 
in a CDM cosmology. He followed the
collapse of gas in a over-dense sphere from high redshift to z
= 0. 
%During the first $\sim$4 Gyrs, the collapse was
%accompanied by a very high star formation rate which essentially
%used up all the available gas, so that after 4 Gyrs (z $\sim$ 3) no new stars
%were formed. 
Star formation was essentially complete in the first 4 Gyrs (by z
$\sim$ 3).  
Kawata \& Gibson (2003) built on this work to simulate 
elliptical galaxies of three different masses
(i.e. 40, 8 and 2 $\times$ 10$^{11}$ M$_{\odot}$).
% and derived
%their properties such as metallicity gradients, central velocity
%dispersions and effective radii. 
They found steeper metallicity
gradients in higher mass galaxies, which was due to the mass
dependence of energy feedback from supernovae. 

Chiosi \& Carraro (2002) used a similar model 
which included supernova feedback 
to describe ``...the monolithic collapse of gas inside
non-rotating, virialized haloes of dark matter...''. Their galaxy
assembly was largely complete by z = 2. They simulated high
and low initial density galaxies over a range of total masses from
10$^8$ to 10$^{13}$ M$_{\odot}$. They also showed that the
resulting metallicity gradients were stronger with increasing galaxy mass. 

Recently, Kobayashi (2004) simulated the formation of elliptical
galaxies in a CDM cosmology, focusing on
internal metallicity gradients. Her models ranged from the assembly of tens
of gas-rich subunits at high redshift (which she denoted as
{\it monolithic collapse}) to the merger of equal mass gas-poor
galaxies at low redshift (called {\it major mergers}). 
She found that galaxies of a given mass 
had steep metallicity gradients if formed by collapse and shallower
gradients (after a few Gyrs had elapsed) if formed by a major merger.
%  -- consistent with previous work that found metallicity gradients to be
%reduced as a result of a merger (e.g. Bekki \& Shioya 1999). 
Thus variations in the formation history of galaxies will tend to
cause scatter in any correlation of gradient with galaxy
mass. 

The galactic wind model of Martinelli \etal (1998) predicts
stronger metallicity gradients in galaxies with larger internal
velocities and hence galaxy mass. Quantative gradient predictions, with
galaxy mass, are not given in their paper.

A gaseous {\it merger} model was presented by Bekki \& Shioya (1999). In
their simulations, two gas-rich disk galaxies of varying mass collide to form an
elliptical galaxy remnant. The gas and stellar masses were set equal so as
to simulate mergers at high redshift. An additional key feature
was the inclusion of supernova feedback. They examined the
metallicity gradients in old merger remnants and found 
depend only weakly on the remnant galaxy mass. 
% which results in
%stronger metallicity gradients than otherwise (although such
%gradients are still some 0.3 dex shallower than those predicted
%by most collapse models).

%The predictions from the simulations described above are shown in
%Fig. 1. 

%\begin{figure}
%\vspace{-4cm}
%\plotone{models.ps}
%\caption[models.ps]{
%Metallicity Gradients with galaxy mass. Each panel shows the
%metallicity gradient and various measures of galaxy mass. We show
%the collapse models of Kawata (2004) by connected open
%squares, the high mass model of Chiosi \& Carraro (2002) by an
%asterisk and the
%monolithically-formed galaxies of Kobayashi (2004) by filled
%squares. The merger model of Bekki \& Shioya (1999) for different
%remnant masses is shown by connected open triangles. 
%}
%\end{figure}

%The figure shows that the steepest gradients are predicted by Kobayashi
%(2004). The gradient predicted by Chiosi \& Carraro
%(2002) is quite consistent with that of Kawata (2004). 
%As expected, the gradients from a merger (Bekki \& Shioya 1999) 
%are shallower still. 

\section{Observational Data}

We now turn to observations of early-type galaxies in the Coma
cluster for which
radial total metallicity gradients have been derived from
single stellar population models. The samples we use are from 
M03 and SB04. 

M03 used the Lick absorption lines (Worthey 1994) of H$\beta$,
Mg$b$, Fe5270 and Fe5335 and the single stellar population models
of Thomas, Maraston \& Bender (2003) to derive metallicity
gradients for early-type Coma galaxies. We take the central velocity
dispersions and effective radii for this sample from Mehlert
\etal (2000).
The spectra of SB04 
cover a large range of Lick lines.
%, with a break
%around 5000\AA ~due to a dichroic. 
They derived metallicity gradients by fitting all available Lick
indices with
%ages and total metallicities using the H$\beta$,
%Mg$b$ and Fe4383 lines (due to the dichroic, the lines of Fe5270 and
%Fe5335 used by Mehlert \etal were not available), and 
the single stellar population models of Vazdekis (1999).  
%Radial gradients were then
%measured from 
%in a similar way to Mehlert \etal (2003), i.e. from
%$\sim$0.1 to $\sim$1r$_e$. 
We also take central velocity dispersions and
effective radii from the work of SB04.

For both samples total K-band apparent magnitudes come from the
2MASS survey, and we calculate absolute magnitudes 
assuming a distance modulus to Coma of 35.0. We have not
corrected the magnitudes for Galactic extinction. This would have
the effect of making each galaxy brighter by $<$0.01$^m$
(Schlegel \etal 1998). We note that the K-band light traces the
underlying mass better, and suffers less from extinction, than
any optical passband. 

Six of the Coma galaxies are in common between the two
studies. We find a mean offset between the gradients of M03 
and SB04 for these six galaxies of 0.25 $\pm$ 0.09 dex per dex. As the
typical gradient in the SB04 study (--0.31) is similar to the
average gradient (--0.30) in the compilation of Kobayashi \& Arimoto
(1999), we elected to subtract 0.25 from the values quoted by
M03. We have implictly assumed that a simple
offset between the two studies is appropriate (with more data, a
more sophisicated analysis could be carried out).
Figure 1 shows the adjusted M03  gradients compared to the
SB04 ones.

\begin{figure*}
%\vspace{-4cm}
\includegraphics[scale=.65,angle=0]{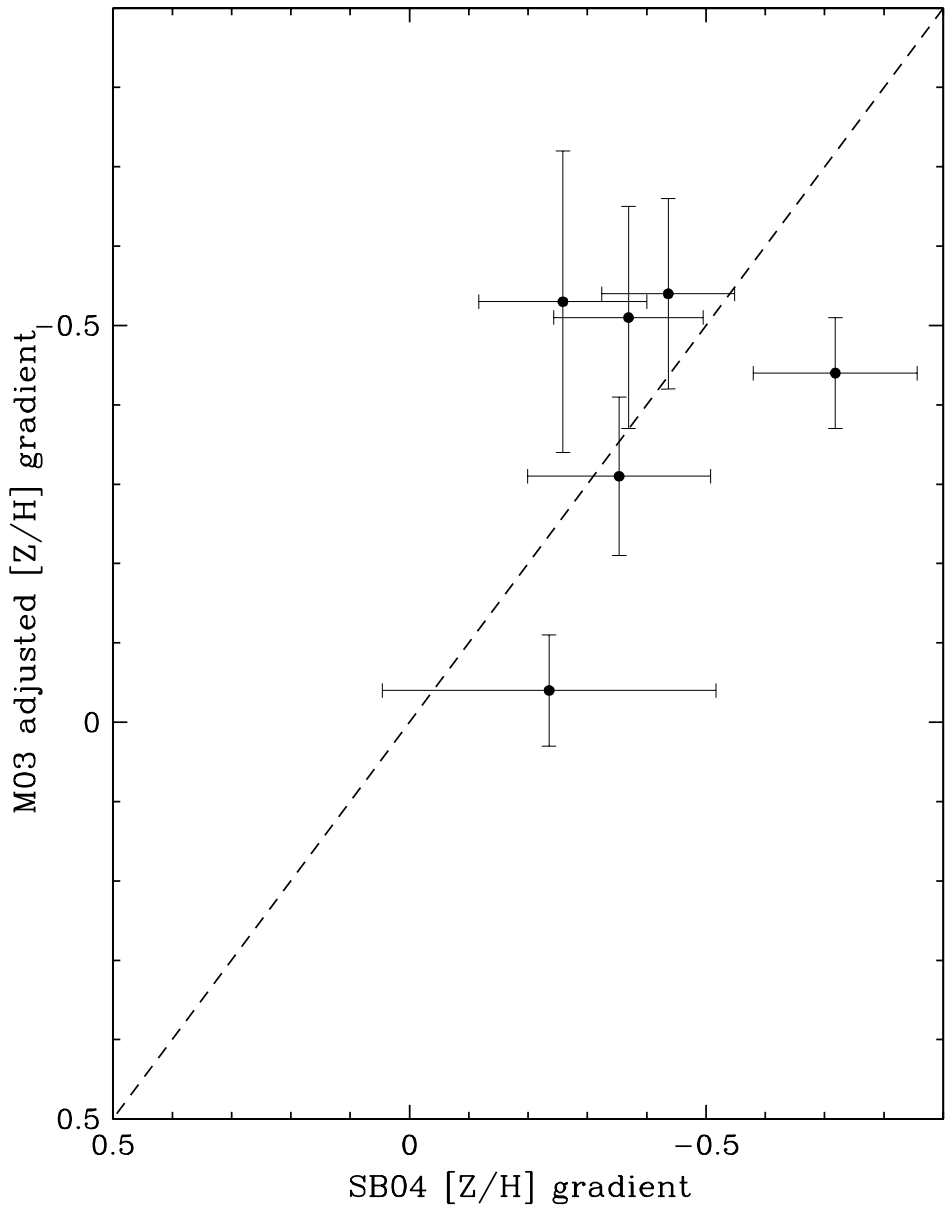}
%\plotone{del.ps}
\caption[compare.ps]{
Comparison of the metallicity gradients. The plot shows the
gradients from Mehlert \etal (2003; M03), after an offset of 0.25 dex
per dex has been applied, against the gradients measured by  S\'anchez-Bl\'azquez
(2004; SB04) for the six Coma cluster
galaxies in common.
}
\end{figure*}

Next we exclude galaxies with central ages younger than 6 Gyrs
from the metallicity gradient analysis. As mentioned earlier,
such young galaxies are prime candidates for a recent gaseous merger
which tends to make luminosity-weighted metallicity gradients
initially much stronger.

We show the metallicity gradients for the 22 M03 and 13
SB04 Coma galaxies with ages $\ge$ 6 Gyrs (our results reported
below do not depend strongly on the exact age limit applied) 
against various proxies
for galaxy mass in Figure 2. Each panel 
shows a different measure of galaxy mass, i.e. absolute K-band magnitude,
central velocity dispersion and the $\kappa_1$ mass parameter  
(defined by Bender, Burstein \& Faber 1992 as $\kappa_1 =
(2 log \sigma_o + log r_e)/\sqrt{2}$). 
Included are the collapse models of Kawata (1999, 2004) and 
the high mass model of Chiosi \& Carraro (2002).
% and the
%monolithically-formed galaxies of Kobayashi (2004). For the
%Kobayashi galaxies we have estimated the gradients from their
%figure 6e and assumed B--V = 0.9. 
We also show the predicted gradients for two merger remnant
ellipticals several Gyrs after the merger event, from the 
simulations of Bekki \& Shioya (1999). The models have been
transformed from the B-band to the K-band assuming B--K = 4 as
appropriate for old stellar populations. 
% for which published gradients are available. 

%The percentage shown 
%is the probability of a correlation from the
%non-parametric Spearman rank test. 

\begin{figure*}
%\vspace{-4cm}
\includegraphics[scale=.65,angle=-90]{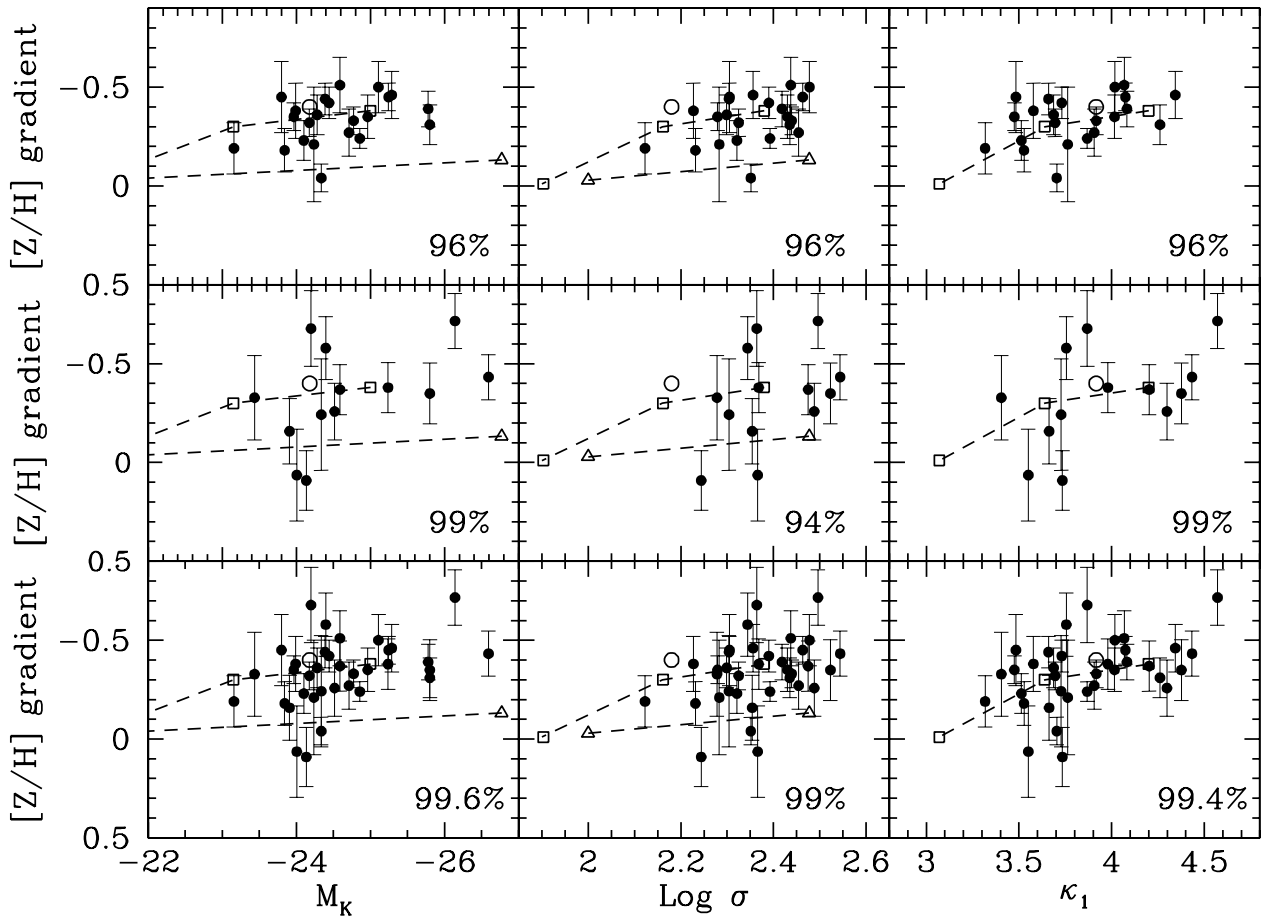}
%\plotone{del.ps}
\caption[panelnew.ps]{
Metallicity gradients with galaxy mass for Coma galaxies. 
Each panel shows the
metallicity gradient and a measure of galaxy mass using the
data for old (age $\ge$ 6 Gyrs) early-type Coma cluster galaxies 
from Mehlert \etal (2003; M03) in the upper panels and from
S\'anchez-Bl\'azquez (2004; SB04) in the middle panels. The lower
panels show the combined data sets. 
Also shown are the three mass models
from the dissipative collapse models of Kawata (2004) as open squares, the
high mass collapse model of Chiosi \& Carraro (2002) as an open
circle, and the merger remnants of Bekki \& Shioya (1999) as open
triangles. The percentage shown 
is the probability of a correlation from the
non-parametric Spearman rank test. Stronger gradients are
generally present
in higher mass galaxies.
}
\end{figure*}

The upper panels show the data of  
M03 (after applying the systematic offset mentioned above), the
middle panels shows the data of SB04 and the lower panel the
simple combination of the two data sets. 
The SB04 gradients reveal a larger scatter (rms $\sim$ 0.5 vs
0.3) than the M03 data.

We have tested the probability of a correlation between the 
metallicity gradient and the measures of galaxy mass in Figure 2 
using a Spearman non-parametric rank test. 
%For all three mass measures of the M03 data, a simple linear
%regression fit suggests that the more
%massive galaxies have stronger metallicity gradients. In
%particular, we find that the metallicity
%gradient correlates most strongly with velocity dispersion with a significance
%of $>$99\% (from a linear correlation coefficient measure). The
%linear fit correlations with M$_B$ and $\kappa_1$ are not statistically significant.
%However, the relation between metallicity gradient and mass may
%not be linear. So we have also 
We find that the metallicity gradient is correlated with the
following probabilities: M$_K$ (96\%), log$\sigma$
(96\%) and $\kappa_1$ (96\%). 
Thus the exclusion of the
young Coma galaxies from the M03 sample reveals
correlations (at the 96\% significance level) 
with mass which were not found in the full sample by M03. 
%We have also divided the sample equally
%into low and high mass subsamples and calculated a mean gradient
%and error on the mean. These values are listed in Table 1. 
% of: --0.32 $\pm$ 0.03 vs. --0.35 $\pm$
%0.04 for M$_B$, --0.29 $\pm$ 0.04 vs. --0.39 $\pm$ 0.03 for
%log$\sigma$ 
%and --0.31 $\pm$ 0.04 vs. --0.37 $\pm$ 0.03 for $\kappa_1$. 
%We find for 
%all three mass measures that the low mass subsample has a shallower
%mean gradient than the high mass galaxies (although the 
%statistically significance is only at the $\sim$2 sigma level). 

%a linear
%fit gives correlations of metallicity gradient with galaxy mass,
%but only the correlation with $\kappa_1$ is statistically
%significant at the $>$90\% level. 
For the SB04 data, 
the Spearman rank test gives correlation probabilities of: M$_K$ (99\%),
log$\sigma$ (94\%) and $\kappa_1$ (99\%). 
%Mean gradients for the low and high mass subsamples are listed in
%Table 1. Again, for all three mass measures the low mass subsample has a shallower
%mean gradient than the high mass galaxies (with 2--3 sigma
%significance).
%$\kappa_1$ is statistically significant.
%have means of: --0.26
%$\pm$ 0.10 vs. --0.39 $\pm$ 0.09 for M$_B$, --0.32 $\pm$ 0.12
%vs. --0.35 $\pm$ 0.09 for log$\sigma$ and --0.19 $\pm$ 0.10
%vs. --0.45 $\pm$ 0.07 for $\kappa_1$. 
For the combined sample of 35 galaxies, the Spearman
probabilities are all improved to: M$_K$ (99.6\%),
log$\sigma$ (99\%) and $\kappa_1$ (99.4\%). 
%The statistical significance of the difference between the 
%low and high mass combined subsamples means are also improved.

We note that the non-parametric Spearman rank test does not take into 
  account the errors of the data points. To address this point, we
  performed 1000 Monte Carlo realizations of the data sample 
in which each point was perturbed randomly with a Gaussian distribution
  of width given by the errors. We ran the Spearman rank order test on 
  each of the 1000 mock data samples and calculated the mode, i.e., the 
  most probable value, of the distribution of
probabilities. These tests reveal that the probability of a
correlation is 99\% or greater for all data sample-mass
combinations (the only exception being the log$\sigma$
correlation for the SB04 sample, with 97\% probability).   
This indicates that the claimed correlations
are real and not due simply to errors in the data. An additional
set of Monte Carlo simulations has shown that the results for the
combined sample is not strongly dependent on the offset applied to
the M03 data. 

Thus we find strong statistical
evidence for a correlation between 
galaxy metallicity gradients and galaxy mass. We also point out
that the correlation is present with galaxy mass parameters 
that are measured completely {\it independently}
of each other (e.g. luminosity and the $\kappa_1$ parameter). This
further strengthens the case for a causal relationship. 

The data scatter fairly evenly about the predictions from the 
dissipative collapse model of Kawata (1999, 2004), and are
generally stronger (more negative) than the  
merger model predictions of Bekki \& Shioya (1999).

\section{Discussion and Conclusions}

Using gradients derived from stellar population models
(rather than simply colours or raw index measures) and by
excluding young galaxies (i.e. allowing
luminosity-weighted gradients
to stablise) we revisit the issue of radial metallicity
gradients in early-type galaxies. 

%In terms of the gradient strengths, the Mehlert \etal
%(2003) Coma galaxy data have values more similar to the 
%gaseous merger model of Bekki \& Shioya (1999) than 
%dissipative collapse models. However, the Mehlert \etal
%measurements appear to be systematically shallower than those
%measured by other authors. 
%The data of SB04 and Proctor (2002) have
%more typical gradient values (see Kobayashi \& Arimoto 1999),
%which show better consistency with
%the dissipative collapse formation models of Chiosi \& Carraro (2002)
%and Kawata (1999, 2004).
%one, as expected at high redshift rather than
%non-dissipational major merging at current epochs. 

Our main result is that 
metallicity gradients for Coma cluster galaxies correlate with
galaxy mass; with a statistical significance of $>$99\% for
the K-band luminosity, 99\% for velocity dispersion and $>$99\% for the
$\kappa_1$ 
mass parameter.     
In general, lower mass galaxies have
shallower gradients. 
Such a trend is consistent with monolithic collapse models 
(Chiosi \& Carraro 2002; Kawata 1999, 2004) which 
invoke gas dissipation and energy ejection (e.g. from
supernova). It is also qualitatively consistent with the galactic
wind models of Martinelli \etal (1998). 
%suggests that some mechanism has acted to reduce the
%gradient, in low mass galaxies, established by gas dissipation. 
The data of SB04, and M03 (after a systematic
offset has been applied), have typical gradients that are more consistent with
the monolithic collapse formation models 
than the gaseous merger model of Bekki \& Shioya (1999). 
This suggests that the dominant formation mechanism for 
old, early-type Coma galaxies is one of monolithic collapse.

To confirm the reported mass trend and further constraint
galaxy formation models, higher signal-to-noise spectra, which
probe to large galactocentric radii, should be obtained. 
Also, samples of low luminosity (M$_K$ $\sim$ --22, M$_B$ $\sim$
--18, $\sigma$ $\sim$ 100 km/s) early-type galaxies at the same
distance (e.g. within a nearby cluster) will be
particularly useful to better define the metallicity
gradient--mass correlation.
% and hence the dominant formation
%mechanism for early-type galaxies. 

\section*{Acknowledgments}

We thank the ARC for financial support. We thank D. Mehlert and
D. Kawata for kindly supplying their data in electronic form. We
also benefited from useful discussions with D. Kawata, and
comments from the anonymous referee. This work
has made use of the NED and LEDA data bases.

\end{document}